# Multi-Objective Operational Optimization of Energy Storage Systems in User-Side Microgrids

Jinzhou Xu, *Student Member, IEEE*, Yuanxin Zhuo, *Student Member, IEEE*, and Paola Tapia, *Member IEEE*

*Abstract*— An operational optimization strategy for microgrid energy storage systems (ESSs) is developed to address practical user-oriented application requirements, and its effectiveness is validated using real-world operational data. First, a fundamental ESS model is established to characterize system dynamics and operational constraints, providing a theoretical basis for optimization. Subsequently, a multi-objective operational optimization framework is formulated to simultaneously minimize electricity cost, reduce carbon emissions, and enhance renewable energy utilization. To ensure computational efficiency and scalability, the commercial optimization solver Gurobi is employed. The proposed strategy is evaluated using actual microgrid operational data, demonstrating that the developed ESS model accurately represents real system constraints. Compared with existing user operational strategies, the proposed approach achieves an average reduction of 13.47% in electricity cost. Moreover, by dynamically adjusting the weighting factors of the multi-objective formulation, the strategy enables flexible operational modes and significantly improves adaptability to varying operating scenarios. In addition, the proposed framework provides decision support for user-side microgrids participating in surplus electricity feed-in policies. The main contribution of this work lies in its user-centric optimization design, which enhances operational flexibility and scenario adaptability through multi-objective weight allocation, offering a practical and scalable solution for real-world microgrid ESS operation.

## I. INTRODUCTION

AS the global energy structure accelerates its transition toward renewable energy, distributed energy resources (DERs) represented by photovoltaic (PV) and wind power are playing an increasingly important role in the construction of modern power systems [1]. However, the inherent intermittency and randomness of renewable generation introduce significant operational challenges, particularly at the user side, where load uncertainty, limited grid support, and flexible operational objectives coexist [2], [3]. Under the guidance of "net-zero" strategic goals, improving renewable energy absorption during renewable energy over-production periods (REOPs), reducing user-side carbon emissions, and enhancing operational economic performance have become urgent and interrelated challenges for user-side microgrids [4].

Energy storage systems (ESSs) are widely recognized as a key enabling technology for balancing power supply and demand, mitigating renewable energy variability, and enhancing renewable energy utilization [5], [6]. Consequently, battery energy storage systems (BESSs) have been increasingly deployed in user-side microgrids to support peak shaving, renewable energy smoothing, and economic operation. Extensive research efforts have been devoted to the control and optimization of BESSs in microgrids [7], [8], [9], [10]. Existing studies on user-side energy storage optimization can be broadly categorized into four main research directions: multi-objective optimal scheduling strategies for energy storage systems [11], [12], optimal configuration, siting, and capacity determination of energy storage systems [13], [14], coordinated optimization of hybrid energy storage systems and multi-energy systems [15], [16], and power distribution and operational control strategies for energy storage systems [17], [18].

From a methodological perspective, various optimization techniques have been proposed to solve microgrid operational optimization problems. Efficient global algorithms for separable quadratic programming and multi-linearly constrained binary quadratic problems have been investigated to improve computational performance and solution quality [19], [20]. Depending on problem formulation characteristics—such as linearity, convexity, and uncertainty modeling—different optimization frameworks have been adopted for microgrid operation and planning [21].

In terms of application focus, existing studies on multi-objective optimal scheduling of ESSs mainly consider economic performance, grid-support capability, and reliability metrics [22]. Research on optimal configuration and siting emphasizes voltage regulation, load fluctuation mitigation, and long-term economic benefits through capacity planning and location optimization [23], [24], [25]. While these works provide valuable theoretical insights, many studies rely on simplified models or predefined operational modes, which limits their adaptability to diverse user-side operating scenarios and evolving policy requirements, particularly under REOP conditions and carbon-reduction constraints.

To address these limitations, this paper develops a comprehensive operational optimization framework for user-side microgrid energy storage systems. First, a fundamental ESS model is constructed to accurately capture operational constraints and system dynamics. On this basis, a multi-

This paragraph of the first footnote will contain the date on which you submitted your paper for review, which is populated by IEEE. It is IEEE style to display support information, including sponsor and financial support acknowledgment, here and not in an acknowledgment section at the end of the article.
  The authors are with the Department of Electrical and Computer Engineering, Universidad Metropolitana de Monterrey, UMM.
  Color versions of one or more of the figures in this article are available online at http://ieeexplore.ieee.org

objective operational optimization strategy is proposed, explicitly incorporating economic cost, carbon emission reduction, and renewable energy absorption objectives. To ensure computational efficiency and scalability for practical deployment, the commercial optimization solver Gurobi is employed. Furthermore, the feasibility and scenario adaptability of the proposed strategy are validated using real-world microgrid operational data. By dynamically adjusting objective weighting factors, the proposed framework enables flexible operational modes tailored to varying user demands and policy environments.

The main contributions of this work can be summarized as follows:
1) A practical and accurate ESS operational model suitable for user-side microgrid applications is established;
2) A user-centric multi-objective optimization strategy integrating economy, carbon emissions, and renewable energy absorption is proposed;
3) A flexible weighting-based mechanism is introduced to enhance scenario adaptability and operational flexibility; and
4) The effectiveness of the proposed strategy is validated using real operational data, demonstrating its applicability for real-world user-side microgrid energy storage operation.

## II. MICROGRID ENERGY STORAGE SYSTEM MODEL

### A. System Architecture

The typical architecture of the user-side microgrid is shown in Fig. 1 [26]. $T_1$ is the distribution transformer for the power grid to supply power to users. The gateway meter for measuring electricity consumption is installed on the low-voltage side of $T_1$. The load L is equivalent to all the loads of users. Users are usually also equipped with photovoltaic (PV) and energy storage (ES). The electrical energy of the load can come from the power grid, photovoltaic, energy storage, or any combination of them.

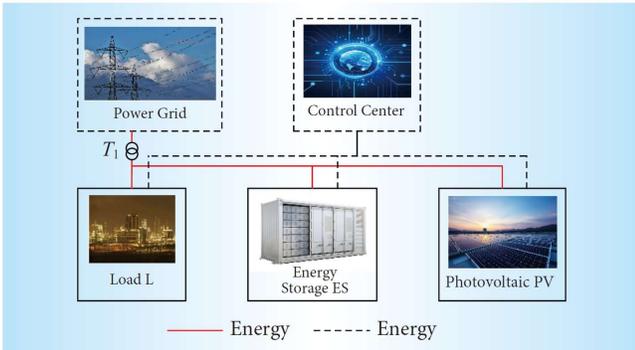

**Fig. 1.** Schematic of typical user-side microgrid.

### B. Energy Storage System Operation Model

The state of the energy storage system at any t time period is one of the three states: charging, discharging, and standby [27].

$$C(t)=\begin{cases} 1 & P_{in}(t)>0 \\ 0 & P_{in}(t)\leq 0 \end{cases} \quad (1)$$

$$D(t)=\begin{cases} 1 & P_{out}(t)>0 \\ 0 & P_{out}(t)\leq 0 \end{cases} \quad (2)$$

$$C(t)+D(t)\leq 1 \quad (3)$$

$$\begin{cases} 0\leq P_{in}(t)\leq P_N^{ES} \\ 0\leq P_{out}(t)\leq P_N^{ES} \end{cases} \quad (4)$$

In the formula: $C(t)$ is the charging state of the energy storage system during the t time period. 1 and 0 respectively indicate whether the energy storage system is in the charging state; $D(t)$ is the discharging state of the energy storage system during the t time period. 1 and 0 respectively indicate whether the energy storage system is in the discharging state; $P_{in}(t)$ and $P_{out}(t)$ are the average charging power and average discharging power of the energy storage system during the t time period respectively; $P_N^{ES}$ is the rated power of the energy storage system, kW.

For ease of understanding, this paper defines $C(t)-D(t)$ as the charging and discharging state in the subsequent expressions. 1 and -1 respectively represent the charging and discharging states, and 0 indicates that the energy storage system is in the standby state.

The state of charge of the energy storage system at any time is equal to the state of charge at the previous time plus (minus) the amount of charge (discharge) during this period [28].

$$\begin{cases} S_{oc}(t)=S_{oc}(t-1)+\dfrac{\eta_c C(t)P_{in}(t)-\dfrac{1}{\eta_D}D(t)P_{out}(t)}{NE} \\ S_{oc,min}\leq S_{oc}(t)\leq S_{oc,max} \end{cases} \quad (5)$$

In the formula: $S_{OC}(t)$ is the state of charge (SOC) at the moment of t, with a value ranging from 0 to 1; $\eta_C$ and $\eta_D$ are the charging and discharging efficiencies of the energy storage system, respectively; E is the rated capacity of the energy storage system, kWh. $0\leq S_{OC,min}\leq S_{OC,max}\leq 1$, which are the SOC limits corresponding to the stop of discharging and charging of the energy storage system, respectively.

When the energy storage system switches from the non-charging (discharging) state to the charging (discharging) state, the number of charging (discharging) times is considered to increase by 1.

$$\begin{cases} N_D(t)=N_D(t-1)+D(t)[1-D(t-1)] \\ N_C(t)=N_C(t-1)+C(t)[1-C(t-1)] \end{cases} \quad (6)$$

In the formula: $N_D(t)$ is the cumulative number of charge-discharge conversions of the energy storage system at the t moment; $N_C(t)$ is the cumulative number of discharge-charge conversions of the energy storage system at the t moment. $N_D(t)$ and $N_C(t)$ are usually set to the same value.

Within two consecutive time periods, when the charging and discharging states of the energy storage system remain unchanged, the value of the output power also remains constant.

$$\begin{cases} C(t-1)C(t)-B_c(t-1)=0 \\ D(t-1)D(t)-B_p(t-1)=0 \\ B_c(t-1)[P_{in}(t)-P_{in}(t-1)]=0 \\ B_p(t-1)[P_{out}(t)-P_{out}(t-1)]=0 \end{cases} \quad (7)$$

In the formula: $B_C(t)$ and $B_D(t)$ are the flag bits for constant-power charging and discharging of the energy storage system, respectively.

## III. Multi-Objective Operation Optimization Strategy for Microgrid Energy Storage Systems

### A. Microgrid Operation Policies

The research object of this paper is the microgrid of industrial and commercial users. In this paper, a two-part electricity price policy is implemented for such users. The electricity bill of users consists of a basic electricity charge and an energy electricity charge. The former is determined by the maximum monthly demand and the basic electricity price; the latter is determined by the real-time electricity consumption and the time-of-use electricity price [29].

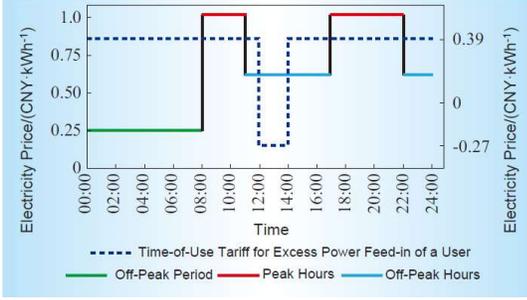

**Fig. 2.** Time-of-use electricity prices for industrial and commercial users.

As shown by the solid line in Fig. 2, the current time-of-use electricity price policy in a certain area consists of three periods: peak, valley, and flat, with prices of 1.0276 $/kWh, 0.2501 $/kWh, and 0.5976 $/kWh respectively. The model of the electricity price policy is shown in Equation (8).

$$p_r(t)=\begin{cases}1.0276 & t\in[8,11)\cup[17,22)\\ 0.5976 & t\in[11,17)\cup[22,24)\\ 0.2501 & t\in[0,8)\end{cases} \quad (8)$$

In the formula: $p_r(t)$ is the electricity price during the t period, in $/kWh.

For the microgrid with an energy storage system and photovoltaic panels as shown in Fig. 1, carbon emissions mainly come from the electricity supplied by the power grid to the microgrid. When calculating the carbon emissions during a certain period, it is usually determined by multiplying the average carbon dioxide emission factor of the local electricity (abbreviated as: carbon emission factor) by the amount of electricity consumed from the power grid during that period. Table 1 shows the carbon emission factors of various prefecture-level cities.

TABLE I
LIST OF AVERAGE ELECTRICITY CARBON EMISSION FACTORS FOR PREFECTURE-LEVEL CITIES

| City code | Electricity carbon factor / (kgCO$_2$·(kWh)$^{-1}$) | City code | Electricity carbon factor /(kgCO$_2$·(kWh)$^{-1}$) |
|---|---|---|---|
| A | 0.642 | H | 0.336 |
| B | 0.677 | J | 0.417 |
| C | 0.689 | K | 0.658 |
| D | 0.653 | L | 0.723 |
| E | 0.631 | M | 0.563 |
| F | 0.623 | N | 0.590 |
| G | 0.171 | | |

The carbon sink price is unified nationwide. As of January 15, 2025, the average carbon sink price in the past 3 months was approximately 0.103 $/kgCO$_2$.

The output of new energy is random, and in many regions, it is regarded as junk electricity. The power grid company will impose economic penalties on the grid - connected new energy during specific REOP periods. Taking the actual users in a certain area as an example, the grid - connected price of surplus electricity from the micro - grid during non - REOP periods is 0.391 $/kWh. To promote new energy consumption, this paper imposes a penalty on users at 0.2703 $/kWh during the REOP period according to the recommended value of experts (in practical applications, it can be determined through negotiation between the power grid and users). As shown by the blue dotted line in Fig. 2, the corresponding new energy consumption electricity price model is

$$p_n(t)=\begin{cases}0.391, & t\notin[t_R,t_S]\\ -0.2703, & t\in[t_R,t_S]\end{cases} \quad (9)$$

In the formula: $p_n(t)$ is the new energy grid - connected penalty factor during the t period, $/kWh; $t_R$ and $t_s$ are the start time and end time of REOP respectively.

### B. Multi - Objective Optimal Operation Model of the Energy Storage System

Formulas (1) to (7) construct the operation model of the energy storage system. On the basis of the above, the energy storage system connected to the micro - grid should also meet the exchange power constraint and power balance constraint.

(1) Exchange power constraint

The power measured by the gateway meter during any t period, whether it is the grid - connected power $P_{dn}(t)$ or the off - grid power $P_{up}(t)$, $P_{dn}(t)$ and $P_{up}(t)$, does not exceed the rated capacity of the distribution transformer, and at any time, at least one of them is 0.

$$\begin{cases}P_{dn}(t)P_{up}(t)=0\\ 0\leq P_{up}(t)\leq P_N^T\\ 0\leq P_{dn}(t)\leq P_N^T\end{cases} \quad (10)$$

In the formula: $P_N^T$ is the rated capacity of the distribution transformer, kVA.

(2) Power balance constraint

For the access point of the energy storage system, the power injected into the node at any t time is equal to the power flowing out of the node.

$$P_{pv}(t)+D(t)P_{out}(t)+P_{dn}(t)=P_{up}(t)+C(t)P_{in}(t)+P_{ld}(t) \quad (11)$$

In the formula: $P_{pv}(t)$ is the average output power of photovoltaics during the t period, kW; $P_{id}(t)$ is the average power of the load during the t period, kW; $P_{in}(t)$ is the average charging power of the energy storage system during the t period, kW.

The user's economic objective is established based on the difference between the electricity purchase cost and electricity sales revenue of the micro - grid in each period. The objective function is as follows

$$F_1=\min\sum_{t=1}^{T}\frac{P_{dn}(t)p_r(t)-P_{up}(t)p_n(t)}{N} \quad (12)$$

In the formula: T represents the number of days considered in the optimization model; N is the number of time periods divided per day. For example, if 15 min is taken as a time interval, then N=96; therefore, N×T represents the total number of time steps in the optimization model. For instance, when there are T=15 days and N=96 (96 15 min periods per day), N×T=1440 indicates that the model covers a total of 15 days of data; $P_{dn}(t)$ is the average power obtained by the microgrid from the power grid during the t period; $P_{up}(t)$ is the average power transmitted by the microgrid to the power grid during the t period, and kW; $p_r(t)$ is the electricity price during the t period, in $/kWh.

Since the carbon emissions at each moment cannot be directly obtained, this paper introduces a real - time carbon emission factor and calculates the carbon emissions through the conversion of the electricity - carbon relationship. In order to put the carbon emission target and the economic target on an equal footing, the carbon emission price is introduced to normalize the carbon emission target. The objective function is as follows

$$F_2 = \min \sum_{t=1}^{T} \frac{P_{dn}(t)c(t)p_c(t)}{N} \quad (13)$$

In the formula: c(t) is the real - time carbon emission factor during the t period, $kgCO_2/kWh$ ; $p_c(t)$ is the equivalent carbon sink price during the t period, in $ /$kgCO_2$ .

2.2.4 New energy consumption target

The purpose of new energy consumption is to make the power generated by new energy be consumed by local loads or energy storage as much as possible, and to minimize the amount of new energy fed into the grid. The proportion of the power output by new energy and energy storage cannot be directly identified at the gateway meter, which needs to be determined by the discharge state of the energy storage and its output power. This paper introduces a new energy grid - connection penalty factor to quantitatively model the indirect economic losses caused by the new energy electricity that is not consumed by the load or energy storage system, and incorporates it as one of the optimization objectives into the multi - objective optimization framework. The objective function is as follows

$$F_3 = \min \sum_{t=1}^{T} \frac{P_{up}(t)[1-D(t)] + [P_{up}(t)-P_{out}(t)]D(t)}{N} p_n(t) \quad (14)$$

In the formula: D(t) is the discharge state of the energy storage system during the t period, where 1 indicates the discharge state and 0 indicates no discharge; $P_{up}(t)$ is the average power transmitted by the microgrid to the power grid during the t period, kW; $P_{out}(t)$ is the output power of the energy storage system during the t period, and kW; $p_n(t)$ is the new energy grid - connection penalty factor during the t period, in $/kWh.

When D(t)=0 , the power measured by the gateway meter is entirely provided by new energy sources. When the unconsumed power of new energy is $P_{up}(t)$; D(t)=1 , the grid-connected power of new energy is equal to the power measured by the gateway meter minus the output power of the energy storage system, and the unconsumed new energy power is $P_{up}(t)-P_{out}(t)$ . In the case where the power generated by new energy is consumed by the energy storage system and then output to the power grid by the latter, this paper defines this part of the electric energy as fully consumed, and the power output from the energy storage system to the power grid is no longer regarded as new energy output.

The comprehensive objective is to perform weighted processing on the economic objective, carbon emission objective, and new energy consumption objective. The user can customize the weights according to the application scenario. The comprehensive objective function is as follows

$$F = \alpha_1 F_1 + \alpha_2 F_2 + \alpha_3 F_3 \quad (15)$$

In the formula: $\alpha_1$ is the economic weight; $\alpha_2$ is the carbon emission weight; $\alpha_3$ is the new energy consumption weight, satisfying $\alpha_1 + \alpha_2 + \alpha_3 = 1$ .

The essence of the operation optimization of the energy storage system is a mixed-integer linear programming (MILP) problem, involving discrete and continuous optimization variables. This paper selects the Gurobi optimizer to solve this model. Gurobi is known for its high performance and strong algorithms, and can efficiently handle large-scale variables and constraints, quickly find the optimal or approximate optimal solution, and narrow the search space through techniques such as the branch and bound method and the cutting plane method.

*C. Microgrid Operation Optimization Strategy*

As shown in Fig. 3, it is the flow chart of the microgrid operation optimization strategy, covering four parts: parameter configuration, photovoltaic prediction, load prediction, and optimization of the energy storage system operation strategy.

All the above processes are completed before 24:00 on the i th day. The output result of the process is the operation strategy of the energy storage system of the microgrid on the i+1 th day. According to the user's operation objectives and policy changes, there may be infinitely many actual optimized operation strategies for the energy storage system on the same day.

IV. CASE STUDY

*A. Case Settings*

This case study analyzes actual microgrid data from City A and City B. City A features industrial users (referred to as User A), with data corresponding to the summer season; City B features commercial users (referred to as User B), with data corresponding to the winter season. The microgrid topologies and time-of-use pricing structures are identical for both cities, as shown in Fig. 1 and Fig.2 respectively. User A has a transformer capacity of 12.5 MVA and a total installed photovoltaic capacity of 6.25 MW. User B has a transformer capacity of 6 MVA and a total installed photovoltaic capacity of 2 MW.

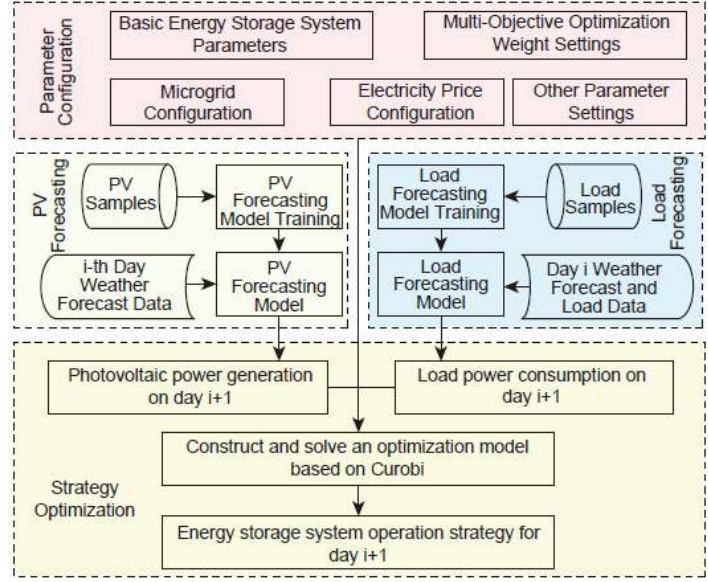

**Fig. 3.** Flow of microgrid operation optimization strategy

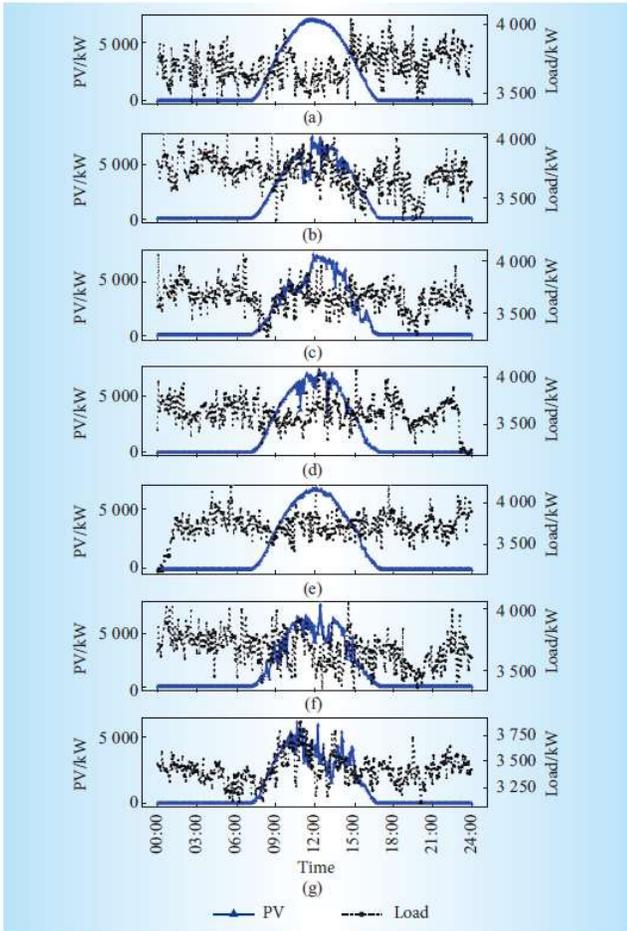

**Fig. 4.** Photovoltaic & load power curves of user A over a week

Fig. 4 shows the photovoltaic generation and load power curves for User A from December 17 to 23, 2024. It is evident that this user's load remains stable year-round at approximately 3.5 MW, while photovoltaic generation is significantly influenced by meteorological factors, exhibiting considerable randomness in output. The microgrid is equipped with one battery energy storage system (BESS) rated at 8 MWh, utilizing lithium iron phosphate (LFP) batteries. User B exhibits similar photovoltaic output characteristics to User A, including common tidal patterns. Their energy storage configuration mirrors that of User A. Relevant electrical parameter configurations are detailed in Table 2.

TABLE II
PARAMETER CONFIGURATION LIST OF BESS

| Project | User A | User B |
| --- | --- | --- |
| Rated capacity $E_N$/kWh | 8000 | 2000 |
| Rated power $P_N^{ES}$/kW | 4000 | 1000 |
| Initial SOC/% | 5 | 5 |
| Lower limit of SOC/% | 5 | 5 |
| Upper limit of SOC | 95 | 95 |
| Charging efficiency $\eta_c$ | 0.88 | 0.85 |
| Discharging efficiency $\eta_n$ | 0.90 | 0.85 |
| Number of charge-discharge cycles $N_c/N_p$ times | 2 | 2 |

The operation strategy of the user's energy storage system is based on the time-of-use electricity price. As shown in Fig. 5, it is the charge-discharge timing diagram of the energy storage system. The charging period of the energy storage system is during the off-peak and flat periods of the time-of-use electricity price; the discharging period of the energy storage system is during the peak period of the time-of-use electricity price to achieve the best peak-valley arbitrage effect.

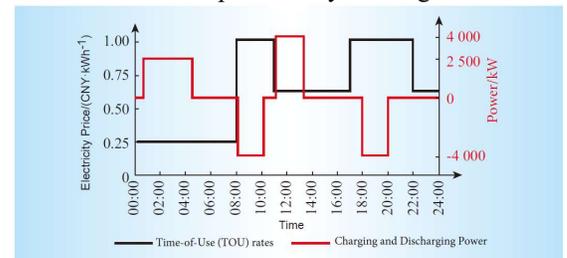

**Fig. 5.** Operation strategy of BESS for user A

*B. Result Analysis*

(1) Number of charge-discharge conversions

Taking the data in Fig. 4(a) as an example, set the number of charge-discharge conversions of the energy storage system to 1 and 2 respectively. As shown in Fig. 6, they are the charge-discharge strategies corresponding to the two fixed values. 1 and -1 represent charging and discharging respectively.

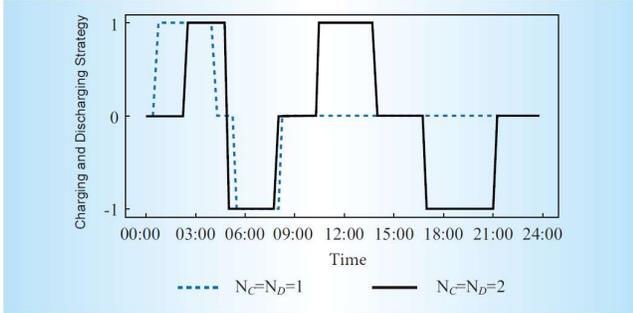

**Fig. 6.** Operation strategy of BESS for a real-world user

By comparing Fig. 5 and Fig. 6, it can be seen that the constraint on the number of charge-discharge conversions of the energy storage system in Equation (6) is effective.

(2) Constant power (dis) charging

Set the energy storage system of user A to operate with constant power (dis) charging, $α_1=1$ , $α_2=α_3=0$ . The corresponding optimized operation strategy of the energy storage system is shown in Fig. 7. The data in each sub - figure of Fig. 7 corresponds to that of Fig. 4 one by one. The red solid line is the charge-discharge power of the energy storage system, and the black dotted line is the SOC of the energy storage system. As can be seen from Fig. 7, when the constant power (dis) charging constraint is enabled, the energy storage system will maintain a constant charge (discharge) power within the same charge (discharge) cycle, which is beneficial to extending the retirement time of the energy storage system and improving the system stability of the microgrid. In the subsequent examples of this chapter, unless otherwise specified, the energy storage system is defaulted to operate in the constant power (dis) charging mode.

The current operation goal of user A only considers economy. The optimized operation strategy of the energy storage system in this paper that only focuses on economy is shown as the red solid line in Fig. 7. Combining with the electricity price policy in Fig. 2, the results of the electricity expenses of users A and B before and after optimization within a week and the percentage of cost reduction are shown in Fig. 8(a) and Fig. 8(b) respectively. The blue and green bar charts represent the users' electricity expenses before and after the optimization of the energy storage system respectively, and the red broken line is the percentage of the users' cost reduction after optimization.

As can be seen from Fig. 8, compared with the original energy storage system operation strategy, the electricity bills of users A and B are significantly reduced after operating with the optimized strategy. The maximum percentage reduction in user A's cost occurred on December 23, with a 14.8482% reduction in electricity bills; the minimum occurred on December 19, with a 12.2516% reduction; and the average reduction in electricity bills within a week was 13.4676%. Except for July 20, the electricity bill reduction on other dates exceeded 15%. The preliminary analysis shows that this is because it was Saturday on that day, and the user's load was too high, which limited the adjustment ability to a certain extent.

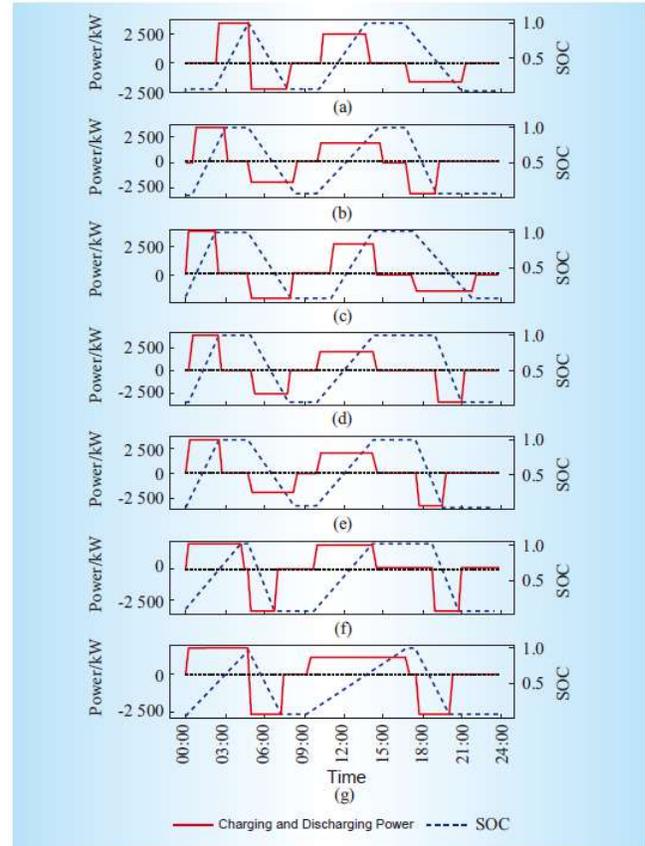

**Fig. 7**. Operation of BESS under constant power (dis) charging

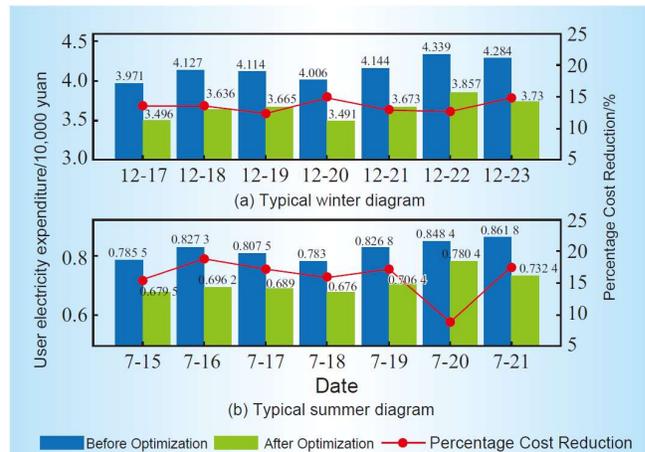

**Fig. 8.** Electricity bills statistics before and after optimization

Taking the data on December 17 in Fig. 4(a) as an example, with 0.1、0.2 and 0.7 as weight factors, six combinations of weight factors for different scenarios $(α_1、α_2、α_3)$ are obtained through permutation and combination. As shown in Table 3, they are the single objective values $(F_1、F_2、F_3)$ corresponding to the six different scenarios and the comprehensive weight values F considering multi-objective optimization, where the $F_1$ column corresponds to the economic objective, which is basically consistent with the data shown in Fig. 8; the $F_1$ and $F_2$ columns correspond to the carbon emission objective and the new energy consumption objective respectively.

It can be seen from Table 3 that when the weight of $\alpha_1$ is the largest (Scenarios 1 and 2), the corresponding value in the $F_1$ column is the smallest; when the weight of $\alpha_2$ is the largest (Scenarios 3 and 4), the corresponding value in the $F_2$ column is the smallest; when the weight of $\alpha_3$ is the largest (Scenarios 5 and 6), the corresponding value in the $F_3$ column is the smallest. Therefore, the proposed optimization strategy for the operation of the microgrid energy storage system is effective for the multi-objective operation of users.

In recent years, the grid-connected capacity of photovoltaics has increased dramatically, and reverse power flow has occurred in more and more distribution transformers. Reasonably setting the moments of $t_R$ and $t_S$ can promote the improvement of the reverse power flow of distribution transformers. Let $t_R=13-i$, $t_S=13+i(i=1,2,3,4)$; the on-grid price of surplus electricity for non-REOP is 0, and the on-grid price of surplus electricity for REOP is $-p_n(t)$; set the weights $\alpha_1=\alpha_2=0$, $\alpha_3=1$. As shown in Table 4 is the grid-connected electricity generation of the microgrid corresponding to different REOP start and end times of the user within a week.

From the data in Table 4, it can be seen that except on December 23rd, the REOP scheme with the minimum grid-connected power of the microgrid within that week is from 11:00 to 15:00. Overall, the results in Column 5 are larger than those in Column 4. When the REOP time window is larger, the suppression effect on the grid-connected power is not necessarily better.

Based on the above idea, this paper increases the REOP with 13:00 as the center and 15 minutes as the step, and sets 12 groups of time windows $[t_R,t_S]$. Using the actual load and photovoltaic data of this user from May 26th to December 31st, 2024, the grid-connected optimization of the microgrid is carried out. Calculate the total grid-connected power corresponding to the 12 time windows every day, and select the time window corresponding to the minimum daily power as the optimal REOP for that day. As shown in Fig. 9, it is the proportion of the optimal REOP corresponding to each time window. Among them, the proportion of the time window [11:15,14:45] is as high as 71.182%, and it is recommended to set it as the REOP period.

TABLE II
MULTI-OBJECTIVE OPTIMIZATION RESULTS UNDER WEIGHTS

| Scenario | $\alpha_i$ | $\alpha_2$ | $\alpha_3$ | $F_1$ / 10,000 $ | $F_2$ / 10,000 $ | $F_3$ / 10,000 $ | F / 10,000 $ |
|---|---|---|---|---|---|---|---|
| 1 | 0.7 | 0.1 | 0.2 | 3.523 | 0.374 | -0.347 | 2.434 |
| 2 | 0.7 | 0.2 | 0.1 | 3.523 | 0.374 | -0.347 | 2.506 |
| 3 | 0.2 | 0.7 | 0.1 | 3.546 | 0.371 | -0.400 | 0.929 |
| 4 | 0.1 | 0.7 | 0.2 | 3.637 | 0.372 | -0.478 | 0.528 |
| 5 | 0.1 | 0.2 | 0.7 | 4.084 | 0.387 | -0.649 | 0.031 |
| 6 | 0.2 | 0.1 | 0.7 | 4.084 | 0.387 | -0.649 | 0.401 |

TABLE II
STATISTICS OF GRID-CONNECTED ELECTRICITY GENERATION UNDER DIFFERENT REOP TIME WINDOWS

| Date | 12:00- 14:00 | 11:00— 15:00 | 10:00- 16:00 | 09:00— 17:00 |
|---|---|---|---|---|
| December 17th | 7.277 3 | 5.8396 | 6.178 2 | 7.7792 |
| December 18th | 5.259 0 | 3.737 5 | 3.9694 | 7.003 6 |
| December 19th | 3.2899 | 2.8414 | 5.9685 | 6.0878 |
| December 20th | 4.038 1 | 3.7808 | 5.400 5 | 6.094 5 |
| December 21st | 4.8450 | 3.5450 | 3.9570 | 6.061 0 |
| December 22nd | 4.891 1 | 2.686 0 | 3.8720 | 4.440 0 |
| December 23rd | 3.5481 | 4.1160 | 4.5489 | 4.0350 |

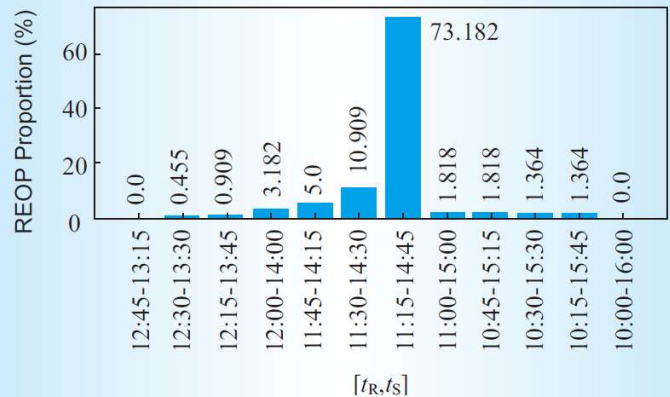

**Fig. 9.** Optimal REOP ratios for different time windows

## VI. Conclusions

This paper develops a comprehensive operational optimization framework for user-side microgrid energy storage systems. First, a fundamental control-oriented model of the energy storage system is established, which accurately captures the operational constraints of practical user-side applications. Based on this model, a multi-objective operational optimization strategy is proposed to address user-oriented requirements, explicitly considering economic performance, carbon emission reduction, and renewable energy utilization. To improve computational efficiency, the commercial optimization solver Gurobi is employed. The proposed strategy is validated using real-world operational data, and the results demonstrate its effectiveness and practical applicability.

The main conclusions can be summarized as follows.

1) The constructed energy storage system control model exhibits constraint consistency with practical user-side energy storage systems, enabling reliable optimization of real-world operational strategies and facilitating the practical deployment of the proposed method.
2) From an economic perspective, the proposed optimization strategy achieves an average reduction of 13.47% in electricity cost compared with existing user operational strategies. This improvement is partially attributed to the accurate prediction of load demand and photovoltaic generation, which enhances optimization performance.
3) The proposed strategy provides decision-making support for surplus electricity feed-in from user-side microgrids, mitigates reverse power flow at distribution transformers, and contributes to improved operational safety and stability of the distribution network.

Overall, the proposed multi-objective operational optimization framework offers a flexible and practical solution for user-side microgrid energy storage operation, supporting cost reduction, low-carbon operation, and enhanced renewable energy integration.